\documentstyle[aps,prl]{revtex}

\begin{document}
\draft
\title{Geodesics in Open Universes}
\author{John D. Barrow${}^{1,2}$ and Janna Levin${}^1$}
\address{$^{1}$Center for Particle Astrophysics,\\
UC Berkeley, 301 Le Conte Hall, Berkeley, CA 94720-7304}
\address{$^{2}$Astronomy Centre, University of Sussex, \\
Brighton BN1 9QH, U.K.}

\preprint{CfPA-97-TH-02}

\twocolumn[
\maketitle
\widetext
\begin{abstract}

We present the geodesics on homogeneous and isotropic negatively 
curved spaces in a simple form suitable for application to 
cosmological problems. We discuss how the 
patterns in the microwave sky of anisotropic 
homogeneous universes can be predicted qualitatively by looking at 
the invariances that generate their three-geometries.

\end{abstract}
\pacs{98.70 Vc, 98.80.Cq, 98.80.Hw}
] \narrowtext
\begin{picture}(0,0)
\put(410,215){{ CfPA-97-TH-02}}
\end{picture} \vspace*{-0.15 in} \setcounter{section}{1}

Metric theories of gravity, of which Einstein's general theory of relativity
is the most elegant known example, couple the material content of the
Universe to its space-time geometry. As a result the paths of light rays
reflect the curvature of space and provide a means to determine the density
and overall curvature of the Universe. Locally, the effect of the Sun's
gravitational field upon grazing light rays and radio waves provides a
powerful test of Einstein's predictions. Globally, we have begun to observe
the effects of gravitational lensing, and we know that varying the density
of the Universe will alter the observed features of microwave background
patterns traced by photons on the sky. Although most versions of the
inflationary universe scenario lead us to expect that the universe will be
expanding very close to the critical divide, with the density parameter
satisfying $\left| \Omega _0-1\right| \sim O(10^{-5})$, there are varieties
of inflation which predict that the universe is significantly open ($\Omega
_0\leq O(0.1)$) \cite{open}. Moreover, the observational evidence stubbornly
refuses to provide a clear endorsement of the $\Omega _0\sim 1$ predictions.
Primordial nucleosynthesis limits the baryon density to fall well short of
this value, and so non-baryonic forms of dark matter must be found in
support of a closure density. For these reasons, the determination of all
the observational differences between flat, closed, and open universes is an
important goal for cosmologists. The most sensitive discriminators promise
to come from the study of null geodesics.

There has been much recent interest in studying the observational signatures
of open universes that possess compact topologies which are produced by the
periodic identification of space coordinates, and of non-compact topologies
that possess identifications in some space directions \cite{{us1},{gs},{bb}}
which permits them to have integrable geodesic motions. Open universes are
interesting candidates for possessing non-trivial topologies because the
curvature of space provides a natural length scale to relate to the scale of
topological identification. There have been several investigations of the
effects of topological identifications on the power spectrum of the
microwave sky and on simulated COBE\ sky maps. These studies enable us to
limit the scale of topological identifications more powerfully than by
searching for multiple images of prominent luminous sources \cite{us1}. In
these studies, the negative curvature of open universes plays an important
role. Geodesic flows on compact negatively-curved spaces are chaotic, as
first noted by Hadamard \cite{had}, and the geodesic flow on a compact
negatively-curved space has become a key paradigm for the identification of
chaotic classical motions as well as in the study of quantum chaos. Some
possible implications for the microwave background have been discussed in
ref. \cite{gurz}. However, the formal characterisation of these mixing flows 
\cite{others} is of little practical utility for the study of geodesics in
the universe. The needs of cosmologists are more specific and in this paper
we aim to provide the geodesics for open universes in simple usable form.
These do not seem to exist in the literature and can be employed to study
the behaviour of the cosmic background radiation in open universes with any
topology. With such applications in mind, we provide a brief analysis of the
geodesics on an expanding negatively-curved space. They are presented in an
immediately accessible form for use in cosmological investigations. Finally,
we provide some new qualitative discussion of the microwave sky patterns in
anisotropic universes. 

\section{Geodesics on a static, negatively curved space}

The most familiar description of the metric in an open universe is in
hyperbolic coordinates 
\begin{equation}
ds^2=-dt^2+dr^2+\sinh ^2r\left( d\theta ^2+\sin ^2\theta d\phi ^2\right) \ \
.  \label{stat}
\end{equation}
However, the geodesics can be found more simply in a coordinate system $%
(x,y,x)$ related to $(r,\theta ,\phi )$ by 
\begin{eqnarray}
e^{-z} &=&\cosh {r}-\sinh {r}\cos {\theta }  \nonumber \\
e^{-z}x &=&\sin {\theta }\cos {\phi }\sinh {r}  \nonumber \\
e^{-z}y &=&\sin {\theta }\sin {\phi }\sinh {r}\ \ .  \label{ge}
\end{eqnarray}
In the $(x,y,z)$ coordinate system the metric is 
\begin{equation}
ds^2=-dt^2+dz^2+e^{-2z}(dx^2+dy^2)\ \ .  \label{one}
\end{equation}
The geodesic equations can be found in the usual way from 
\begin{equation}
{\frac{d^2x^\mu }{d\lambda ^2}}+\Gamma _{\alpha \beta }^\mu {\frac{dx^\alpha 
}{d\lambda }}{\frac{dz^\beta }{d\lambda }}=0\ \ .
\end{equation}
but it is more efficient to introduce the Lagrangian 
\begin{equation}
L={\frac 12}\left[ -t^{\prime 2}+z^{\prime 2}+e^{-2z}\left( x^{\prime
2}+y^{\prime 2}\right) \right]
\end{equation}
where ${}^\prime={d/d\lambda }$. The equations of motion can be found from 
\begin{equation}
\Pi _q^{\prime }-{\frac{\partial L}{\partial q}}=0
\end{equation}
where $\Pi _q=\partial L/\partial q^\prime$ is the momentum conjugate to
coordinate $q$.

Since the Lagrangian is independent of $(t,x,y)$, the corresponding
conjugate momenta are conserved, giving 
\begin{eqnarray}
\Pi _t=-t^{\prime } &=&-E_{{\rm i}}  \nonumber \\
\Pi _x=e^{-2z}x^\prime &=&\Pi _{x{\rm i}}  \nonumber \\
\Pi _y=e^{-2z}y^\prime &=&\Pi _{y{\rm i}}  \label{piz}
\end{eqnarray}
where $E_{{\rm i}},\Pi _{x{\rm i}},\Pi _{y{\rm i}}$ are constants of the
motion. The last of these equations is 
\begin{equation}
z^{\prime \prime }+e^{-2z}(x^{\prime 2}+y^{\prime 2})=0\ \ .
\end{equation}
This second-order equation can be reduced to first order by exploiting the
invariance of the length element. Choosing the affine parameter to be proper
time $\eta $, $ds^2/d\eta ^2=\alpha $ ($\alpha =0$ for photons and $\alpha
=-1$ for massive particles), we have (\ref{one}) 
\begin{equation}
\alpha =-E_{{\rm i}}^2+z^{\prime 2}+e^{-2z}\left( x^{\prime 2}+y^{\prime
2}\right) \ \ .
\end{equation}
Substituting the solutions for ${x^{\prime }}$ and ${y^{\prime }}$ and
defining $W_{{\rm i}}^2=\Pi _{x{\rm i}}^2+\Pi _{y{\rm i}}^2$, we can solve
this for ${z^{\prime }}$ to find 
\begin{equation}
{z^{\prime }}=\pm \left[ (\alpha +E_{{\rm i}}^2)-e^{2z}W_{{\rm i}}^2\right]
^{1/2}\ \ .  \label{velx}
\end{equation}
The system has been reduced to first-order equations. Notice that $z^{\prime
\prime }\le 0$ always. It follows that if ${z^{\prime }}>0$, then $z$ will
reach a maximum and then the geodesic reverses direction.

Integrating the $z$ equation, we have 
\begin{equation}
\int_{z_{{\rm i}}}^z{\frac{dz}{\left[ (\alpha +E_{{\rm i}}^2)-e^{2z}W_{{\rm i%
}}^2\right] ^{1/2}}}=\pm (\eta -\eta _{{\rm i}}).\ \   \label{left}
\end{equation}
This completes the solution. Let the initial time be $\eta _{{\rm i}}=0$.
Firstly, if $W_{{\rm i}}^2=0$, the trajectories are simple lines, 
\begin{equation}
{W_{{\rm i}}^2=0\ \ \ \ \ z=z_{{\rm i}}\pm \gamma \eta ,}
\end{equation}
with $\gamma =(\alpha +E_{{\rm i}}^2)^{1/2}$. If $W_{{\rm i}}\ne 0$ we can
integrate (\ref{left}) and then (\ref{ge}) to obtain 
\begin{eqnarray}
W_{{\rm i}}^2e^{2z} &=&\gamma ^2{\frac 1{\cosh ^2(\gamma \eta \mp \beta _{%
{\rm i}})},}  \nonumber \\
x &=&x_{{\rm i}}+{\frac{\Pi _{x{\rm i}}}{W_{{\rm i}}^2}}\gamma \left[ \tanh
(\gamma \eta \mp \beta _{{\rm i}})\pm \tanh (\beta _{{\rm i}})\right] , 
\nonumber \\
y &=&y_{{\rm i}}+{\frac{\Pi _{y{\rm i}}}{W_{{\rm i}}^2}}\gamma \left[ \tanh
(\gamma \eta \mp \beta _{{\rm i}})\pm \tanh (\beta _{{\rm i}})\right] .
\label{oneway}
\end{eqnarray}
The last constant of integration that appears here is 
\begin{equation}
\tanh \beta _i=\sqrt{\gamma ^2-e^{2z{\rm i}}W_{{\rm i}}^2}/\gamma \ \ .
\end{equation}
Notice that, unless a particle is shot directly along the axis, it will
never reach $z=+\infty $. The constant, which is proportional to $\tanh
(\beta _{{\rm i}}),$ could be absorbed into $x_{{\rm i}}$, but then $x_{{\rm %
i}}$ and would not correspond to the value $x(\eta _{{\rm i}}=0)$. Rewriting
these in another way, that is useful when tracing geodesics, gives 
\begin{eqnarray}
e^{-z} &=&e^{-z_{{\rm i}}}\left[ \gamma \cosh (\gamma \eta )\mp \sinh
(\gamma \eta )\sqrt{\gamma ^2-e^{2z_{{\rm i}}}W_{{\rm i}}^2}\right] 
\nonumber \\
x &=&x_{{\rm i}}+{\Pi _{x{\rm i}}e^{2z_{{\rm i}}}}\gamma ^2\left[ \frac{%
\tanh (\gamma \eta )}{\gamma \mp \tanh (\gamma \eta )\sqrt{\gamma ^2-e^{2z_{%
{\rm i}}}W_{{\rm i}}^2}}\right]  \nonumber \\
y &=&y_{{\rm i}}+{\Pi _{y{\rm i}}e^{2z_{{\rm i}}}}\gamma ^2\left[ \frac{%
\tanh (\gamma \eta )}{\gamma \mp \tanh (\gamma \eta )\sqrt{\gamma ^2-e^{2z_{%
{\rm i}}}W_{{\rm i}}^2}}\right]  \label{another}
\end{eqnarray}
where 
\begin{equation}
{\frac 1{\cosh ^2{\beta _{{\rm i}}}W_{{\rm i}}^2}}=\exp (2z_{{\rm i}})\ \ .
\end{equation}

We can write the path parametrically as 
\begin{equation}
e^{2z}={\frac{\gamma ^2}{W_{{\rm i}}^2}}-(x-\hat x_{{\rm i}})^2-(y-\hat y_{%
{\rm i}})^2
\end{equation}
where $\hat x_{{\rm i}}=x_{{\rm i}}\pm (\Pi _{x{\rm i}}/W_{{\rm i}}^2)\gamma
\tanh (\beta _{{\rm i}})$ and $\hat y_{{\rm i}}$ is defined analogously.

These trajectories have some very odd properties. As already mentioned, the
trajectories never reach $z=+\infty $. Any photon travelling along
increasing $z$ eventually hits a maximum and then wraps back. As it moves
toward $z=-\infty $, the velocities along $x$ and $y$ fall (eqns (\ref{piz}%
)) so that they never reach infinite values of $x$ or $y$.


\subsection{Geodesic flows in an expanding universe}

When the expansion of space is included, the full metric becomes that of the
open Friedmann universe, 
\begin{equation}
ds^2=-dt^2+a^2(t)\left[ dz^2+e^{-2z}(dx^2+dy^2)\right] \ \ .
\end{equation}
The {\it null} geodesic equations are 
\begin{equation}
\ddot t+H\dot t^2=0  \label{time}
\end{equation}
\begin{equation}
\ddot z+e^{-2z}\left( \dot x^2+\dot y^2\right) +2{\cal H}\dot z=0
\end{equation}
\begin{equation}
\ddot x-2\dot z\dot x+2{\cal H}\dot x=0
\end{equation}
\begin{equation}
\ddot y-2\dot z\dot y+2{\cal H}\dot y=0\ \ ,
\end{equation}
where an overdot denotes $d/d\lambda $ and ${\cal H}=H\dot t=d\ln {a}%
/d\lambda $. Notice from $(\ref{time})$ that 
\begin{equation}
\dot t={\frac 1a}\ \ .
\end{equation}
so $d\lambda =adt$.

All of the results of the previous section can quickly be adapted to the
case with expansion if a time coordinate is chosen astutely. Now let $\prime
{}=d/d\eta $. Then the geodesic equations become 
\begin{equation}
\dot \eta ^2\left[ z^{\prime \prime }+e^{-2z}({x^{\prime }}^2+{y^{\prime }}%
^2)\right] +z^\prime\left[ \ddot \eta +2{\cal H}\dot \eta \right] =0
\end{equation}
\begin{equation}
\dot \eta ^2\left[ x^{\prime \prime }-2{y^{\prime }}{x^{\prime }}\right] +{%
x^{\prime }}\left[ \ddot \eta +2{\cal H}\dot \eta \right] =0
\end{equation}
\begin{equation}
\dot \eta ^2\left[ y^{\prime \prime }-2{z^{\prime }}{y^{\prime }}\right]
+y^\prime\left[ \ddot \eta +2{\cal H}\dot \eta \right] =0\ \ .
\end{equation}
If we choose the coordinate $\eta $ such that 
\begin{equation}
\ddot \eta +2{\cal H}\dot \eta =0,
\end{equation}
then it follows that the geodesic equations become 
\begin{eqnarray}
z^{\prime \prime }+e^{-2z}\left( {x^{\prime }}^2+{y^{\prime }}^2\right) &=&0
\nonumber \\
x^{\prime \prime }-2{z^{\prime }}{x^{\prime }} &=&0 \\
y^{\prime \prime }-2{z^{\prime }}{y^{\prime }} &=&0  \nonumber
\end{eqnarray}
which is precisely the same as those on a static, negatively curved
hypersurface. The solutions are then the same as equations (\ref{oneway})
(or equivalently (\ref{another})) with $\gamma =1$ for photons and 
\begin{equation}
\eta =\int {\frac{d\lambda }{a^2}}=\int {\frac{dt}{a(t)}}\ \ 
\end{equation}
is the usual conformal time.

For timelike geodesics, the motion can again be projected onto a static
hypersurface with a new affine parameter. Again, we recover the geodesic
flows of the previous section with $\alpha =-1$ except the time parameter
for massive particles is not conformal time but rather \cite{lock} 
\begin{equation}
\eta =\int {\frac{dt}{a(v_o+a^2)^{1/2}}}\ \ 
\end{equation}
and $v_o=av(t)/\sqrt{1-v^2}$ and $v^2(t)=g_{ij}\dot x^i\dot x^j$.

\section{Tracing geodesics}

The microwave background provides a sensitive probe of the curvature of the
universe. If we locate the Earth (or our near-Earth satellite) at the origin
of the coordinate system, then only photons from the surface of last scatter
which travel along radial geodesics will be observed. The geodesic equations
are then greatly simplified with respect to the direction of observation on
the sky.

The photons seen in the sky today can be traced backwards to locate their
point of origin. We have six unknowns: 
\begin{eqnarray}
\ z_{{\rm i}}&<=>\beta _{{\rm i}}\quad \quad &z_{{\rm i}}^{\prime }
\nonumber \\
x_{{\rm i}} &&\Pi _{x{\rm i}}  \nonumber \\
y_{{\rm i}} &&\Pi _{y{\rm i}}\ \ .
\end{eqnarray}
The magnitude of $z_{\rm i}^\prime $ is fixed by eqn (\ref{velx}) once the 
other initial coordinates are specified.
The choice $z_{\rm i}^\prime $ corresponds then to a choice of the sign in 
eqn (\ref{velx}).
There are two sets of boundary conditions: ($i$) The photon position vector
is at the origin, and ($ii$) The velocity vector is opposite to the unit
vector pointing in the direction of observation. In other words, 
\begin{eqnarray}
(i)\quad \vec x &=&\vec x_o=\vec 0  \nonumber \\
(ii)\quad \hat v &=&-\hat n(\theta ,\phi )\ \ .
\end{eqnarray}
Boundary condition $(i)$ places the Earth at the origin of the coordinate
system. Using the geodesic solutions, we have 
\begin{eqnarray}
W_{{\rm i}}^2 &=&{\frac 1{\cosh ^2(\eta _o\mp \beta _{{\rm i}})},}  \nonumber
\\
x_{{\rm i}} &=&-{\frac{\Pi _{x{\rm i}}}{W_{{\rm i}}^2}}\left[ \tanh (\eta
_o\mp \beta _{{\rm i}})\pm \tanh (\beta _{{\rm i}})\right] ,  \nonumber \\
y_{{\rm i}} &=&-{\frac{\Pi _{y{\rm i}}}{W_{{\rm i}}^2}}\left[ \tanh (\eta
_o\mp \beta _{{\rm i}})\pm \tanh (\beta _{{\rm i}})\right] \ \ .  \label{xc}
\end{eqnarray}
When evaluated at the origin, the geodesic equations relate the components
of the velocity vector today to their initial values: 
\begin{eqnarray}
z_o^{\prime 2} &=&1-W_{{\rm i}}^2  \nonumber \\
x_o^{\prime } &=&\Pi _{x{\rm i}}  \nonumber \\
y_o^{\prime } &=&\Pi _{y{\rm i}}\ \ .
\end{eqnarray}
This velocity vector is normalized to $1$ as it must be for photons. Using 
\begin{equation}
\hat n=\hat r=\sin \theta \cos \phi \hat {{\rm i}}+\sin \theta \sin \phi 
\hat {{\rm j}}+\cos \theta \hat {{\rm k}}\ ,\ 
\end{equation}
we can rotate this into the $(x,y,z)$ coordinate system at the origin to
find 
\begin{eqnarray}
\hat z<=> &&\hat {{\rm k}}  \nonumber \\
\hat x<=> &&\hat {{\rm j}}  \nonumber \\
\hat y<=> &&\hat {{\rm i}}\ \ .
\end{eqnarray}
Condition $(ii)$ then gives 
\begin{eqnarray}
z_o^{\prime } &=&-\cos \theta  \nonumber \\
x_o^{\prime }=\Pi _{x{\rm i}} &=&-\sin \theta \cos \phi  \nonumber \\
y_o^{\prime }=\Pi _{y{\rm i}} &=&-\sin \theta \sin \phi \ \ .  \label{vc}
\end{eqnarray}
\mbox{$>$}
It follows that $W_{{\rm i}}^2=\sin ^2\theta $. Also, taking the derivative
of the geodesic equation, 
\begin{equation}
z_o^{\prime }=-\tanh (\eta _o\mp \beta _{{\rm i}})=-\cos \theta ,\ \ 
\end{equation}
from which it follows that 
\begin{equation}
\mp \tanh (\beta _{{\rm i}})={\frac{\cos \theta -\tanh (\eta _o)}{1-\tanh
(\eta _o)\cos \theta }}\ \ 
\end{equation}
$\beta _{{\rm i}}=\pm (\eta _o-{\rm arctanh}(\cos \theta ))$. Putting (\ref
{vc}) into (\ref{xc}) we find 
\begin{eqnarray}
e^{-z_{{\rm i}}} &=&{\ \cosh (\eta _o)-\sinh (\eta _o)\cos \theta } 
\nonumber \\
e^{-z_{{\rm i}}}x_{{\rm i}} &=&{\sin \theta \cos \phi }\sinh (\eta _o) 
\nonumber \\
e^{-z_{{\rm i}}}yi &=&{\sin \theta \sin \phi }\sinh (\eta _o)\ \ .
\label{xca}
\end{eqnarray}
These are all radial geodesics in {\it $(r,\theta ,\phi )$; }that is,{\it \ $%
r=\eta _o-\eta $.} The integrated Sachs-Wolfe effect considers only radial
geodesics; but processes such as gravitational lensing, that can deflect a
photon into the line of sight, would draw from the more general pool of
non-radially directed photons.

\section{Local and Global anisotropy}

The full geodesics on a universe of negative curvature have been obtained in
an explicit form most accessible to cosmologists. One arena of renewed
interest where the full geodesics may be needed is the case of a small
universe. Negatively curved spacetimes can be made small and finite through
topological identifications. A small universe could be witnessed with
periodic effects or by features in the power spectrum of the microwave
background. Topology induces global anisotropy even when the underlying
space is locally isotropic. Local anisotropy is also possible in the absence
of topological identifications. When homogeneous anisotropies are present,
either in the form of shear or rotation in the expansion of the universe, or
possibly also in the three-curvature of space, there are only a finite
number of homogeneous spaces which can provide an exact description of the
geometry of space. These anisotropic spaces were first classified by Bianchi 
\cite{bian}. They were introduced into cosmology by Taub \cite{taub}, and
presented in the most efficient manner by Ellis and MacCallum \cite{mac}.
Since the microwave sky is currently the most significant historical record
of the primordial radiation, it is instructive to show how the anisotropic
sky patterns created by these different anisotropic universes can be
predicted just from a knowledge of the group invariances that generate the
homogeneous geometries and their geodesic flows.

In order to determine the detailed sky patterns permitted by the Bianchi
geometries it is necessary to solve for the evolution of the geodesics on
the anisotropic cosmological models either exactly or approximately (in the
case of small anisotropy); see for example refs. \cite
{{Nov},{hawk},{CollHaw},{BJS1},{BJS2}}. Again, the most unusual features
arise in open universes. The basic quadrupole pattern arises in the simplest
(flat) Bianchi type I universe with zero curvature. The addition of negative
curvature focuses this quadrupole into a small hotspot on the sky (there is
a preferred direction because there is a direction of lowest and highest
expansion rate) in type V universes, which still possess isotropic
3-curvature. If anisotropic curvature is added then we reach the most
general class of anisotropic homogeneous spaces and a spiralling of the
geodesics is added to the quadrupole or focused quadrupole in the flat or
open type VII universes. These particular models have been studied in the
past by linearizing the geodesic equations about the isotropic solutions in
which the temperature anisotropy of the microwave background is zero \cite
{{Nov},{hawk},{CollHaw},{dzn},{BJS1},{BJS2},{JDB1},{bunn}}. They describe
the most general anisotropic distortions of flat and open Friedmann
universes. However, it is also possible to predict the geometric sky
patterns expected in the different Bianchi type universes by simply noting
the nature of the groups of motions which define each homogeneous space.

The Bianchi classification of spatially homogeneous anisotropic universes is
based on the geometric classification of 3-parameter Lie groups. The action
of these groups on the spacelike hypersurfaces of constant time in these
universes can be prescribed by three transformations of cartesian
coordinates $(x,y,z)$. Each model possesses two simple translations in the $%
x-y$ plane, with generators $\partial /\partial x$ and $\partial /\partial y,
$ together with a more complicated motion out of this plane which is
different for each group type. If it is considered as a flow from the $z=0$
plane to some other plane, $z=\alpha =$ constant, then the nature of this
flow tells us qualitatively what the microwave background anisotropy pattern
will look like. In the simplest flat universe of Bianchi type I the $z-$flow
is uniform and just maps $(x,y)\rightarrow (x,y).$ This corresponds to a
pure quadrupole geodesic temperature anisotropy pattern. In the open Bianchi
type V universe the $z-$flow is a pure dilation and maps $(x,y)\rightarrow
e^\alpha (x,y),$ with $\alpha $ constant$.$ This dilation describes the
hotspot created by the focusing of the quadrupole pattern in open
anisotropic universes. The most general non-compact homogeneous universes
containing the Friedmann models are of Bianchi type VII$_h$, which contain
the flat Friedmann universes when $h=0$ and the open Friedmann universes
when $h\neq 0$. In type VII$_h$ the $z-$flow is a rotation plus a dilation:
circles of radius $r=(x^2+y^2)^{1/2}=1$ are mapped into circles of radius $%
r=e^{\alpha \sqrt{h}}$ and rotated by a constant angle $\alpha .$
Observationally, this corresponds to the geodesics producing a focusing of
the basic quadrupole (as in type V) with a superimposed spiral twist. In
Bianchi type VII$_0$ there is simply a spiral added to the underlying
quadrupole with no focusing because the 3-geometry is flat. The $z-$flow for
the closed universes of Bianchi type IX, which contain the closed Friedmann
models as special cases, is more complicated. The $z-$plane action
corresponds to the following $SO(3)$ invariant motion in polar coordinates
based on the $x$-axis, in the region $r<2\pi ,$\cite{sik}: 
\begin{eqnarray*}
\{r,\theta ,\phi \} &=&\{\cos ^{-1}[(1-\beta ^2)^{1/2}\cos (\alpha /2)], \\
&&\sin ^{-1}\beta [1-(1-\beta ^2)\cos ^2(\alpha /2)]^{-1/2},\ \frac 12\alpha
+\phi _0\}
\end{eqnarray*}
Sky patterns in other, less familiar, Bianchi types can be generated in a
similar fashion, if required. Thus, type VI$_h$ is generated by a $z-$flow
that combines a shear with a dilation: the hyperbolae $x^2-y^2=A^2$ are
mapped into hyperbolae which are rotated by a hyperbolic angle $\beta $ into 
$x^2-y^2=A^2e^{2\beta \sqrt{-h}}.\ $

\begin{center}
\_\_\_\_\_\_\_\_\_\_\_\_\_\_\_\_\_\_\_\_\_\_\_\_\_\_\_
\end{center}

\vskip10truept

We extend special thanks to P. Ferreira and J. Silk for discussions about
these ideas. JDB is supported by the PPARC. JJL is supported in part by
a President's Postdoctoral Fellowship.

\end{document}